\documentclass[twoside]{PoS}

\PoS{PoS(HEP2005)246}

\usepackage{graphicx,cite,amssymb,epsfig,float,psfrag,multirow,amsmath,subfigure}

\catcode`@=11

\def\@citex[#1]#2{\if@filesw\immediate\write\@auxout{\string\citation{#2}}\fi
  \def\@citea{}\@cite{\@for\@citeb:=#2\do
    {\@citea\def\@citea{,\penalty\@m}\@ifundefined
      {b@\@citeb}{{\bf ?}\@warning
       {Citation `\@citeb' on page \thepage \space undefined}}%
\hbox{\csname b@\@citeb\endcsname}}}{#1}}

\def\citer{\@ifnextchar [{\@tempswatrue\@citexr}{\@tempswafalse\@citexr[]}}

\def\@citexr[#1]#2{\if@filesw\immediate\write\@auxout{\string\citation{#2}}\fi
  \def\@citea{}\@cite{\@for\@citeb:=#2\do
    {\@citea\def\@citea{--\penalty\@m}\@ifundefined
       {b@\@citeb}{{\bf ?}\@warning
       {Citation `\@citeb' on page \thepage \space undefined}}%
\hbox{\csname b@\@citeb\endcsname}}}{#1}}

\def\citer{\@ifnextchar [{\@tempswatrue\@citexr}{\@tempswafalse\@citexr[]}}
\catcode`@=12

\newcommand{\beq}{\begin{equation}}
\newcommand{\eeq}{\end{equation}}
\def\Bbar{\overline{B}^0}

\def\nn{\nonumber}

\def\rt{{r_t}}
\def\dt{{r_t}}
\def\phips{{\phi_+}}
\def\phims{{\phi_-}}

\def\etab{{\bar{\eta}}}

\def\Spi{S_{\pi\pi}}
\def\Srho{S_{\rho\rho}}

\def\gsim{ {\ \lower-1.2pt\vbox{\hbox{\rlap{$>$}\lower5pt\vbox{\hbox{$\sim$}}}}\ } }
\title{The Unitarity Triangle through $B_d\to \rho^{\pm} \pi^{\mp}$ decays
}

\ShortTitle{The Unitarity Triangle through $B\to \rho^{\pm} \pi^{\mp}$ decays}

\vspace*{-0.4cm}
\author{\speaker{A.~Salim Safir}\\
     CERN, Department of Physics, Theory Unit, CH-1211 Geneva 23, Switzerland\\
        E-mail: \email{safir@mail.cern.ch}}

\abstract{
We analyze the impact of the  CP-violating observables in the
$B_d\to \rho^{\pm} \pi^{\mp}$ system, combined with the precise
measurement of $\sin2\beta$ , in the extraction of the CKM matrix.
We explore two strategies for determining  the Unitarity Triangle in these
modes.
Computing the penguin parameters $(r_{\pm}, \phi_{\pm})$ and the ratio of two
trees $(r_{t}, \phi_{t})$ within QCD factorization yields a precise
determination of $(\bar\rho, \bar\eta)$, reflected by a weak dependence on
$\phi_{\pm}$, which is shown to be a second order effect, as
in the $B_d \to \pi^+ \pi^-$ system. Moreover, we find that the dependence on
penguin amplitudes $r_{\pm}$ in $B_d\to \rho^{\pm} \pi^{\mp}$ is less pronounced
than in the $B_d \to \pi^+ \pi^-$ case, since penguin contributions
$r_{\pm}\approx r_{\pi\pi}/3$, implying an important simplification in our
analysis. Independent experimental tests of the factorization framework are
proposed and discussed, using $B_d \to \pi^+ \pi^-$ and $B_d \to \rho^{\pm}
\rho^{\mp}$ modes.}

\FullConference{International Europhysics Conference on High Energy Physics\\

                 July 21st - 27th 2005\\

                 Lisboa, Portugal}

\begin{document}

One of the most relevant challenges for the $B$-factories is the determination 
of the three angles of the Unitarity Triangle (UT) of the 
Cabibbo-Kobayashi-Maskawa (CKM) matrix \cite{CKM-KM}. To date, besides 
the precisely measured angle $\beta$, from the ``gold-plated'' mode 
$B_d \to J/\psi K_S$, the extraction of the two remaining 
angles, namely $\alpha$ and $\gamma$, is obscured by our lack of knowledge
about the hadronic dynamics inside mesons. Although their extraction is
mainly limited theoretically by the so-called penguin pollution,  CP 
violation in the charmless $B$ decays, such as $B_d\to \pi\pi, \pi \rho$ 
and similar modes, could be of great help in their extractions.

%


In this work, we propose a transparent analysis of exploring the UT 
through the CP violation in $B_d\to \rho^{\mp}\pi^{\pm}$, combined with 
the ``gold-plated'' mode $B_d \to J/\psi K_S$. 
Contrary to the $B_d\to \pi^-\pi^+$ mode $B_d\to \rho^{\mp}\pi^{\pm}$ 
exhibits 
two transition amplitudes, namely 
$A_{\pm} \equiv A(B^0\to \pi^{\pm}\rho^{\mp})=
T_{\pm}\, \bigg(e^{i\gamma} + \frac{P_{\pm}}{T_{\pm}}\bigg)$, where
$T_{\pm}$ and $P_{\pm}$ are respectively the corresponding tree and penguin 
amplitude. Since the determination of the penguin-to-tree amplitude
is relevant for our analysis, we use QCD factorization (QCDF) \cite{BBNS1} 
for their estimate.

The time-dependent decay rates for $B_d\to\rho^\pm\pi^\mp$ decays
are defined by 6 Observables $C,~\Delta C$, $S,~\Delta S$,
~$\Gamma^{\rho \pi}$ and ${\cal A}^{\rho \pi}$, insufficient input to predict
model independently the 8 theoretical parameters, namely 
7 hadronic parameters $|T_\pm|, |P_\pm|, \phi_\pm, \phi_t $ and one 
weak phase $\gamma$ related to these modes. To disentangle the CKM 
dependence from the hadronic parameters, we write the penguin-to-tree ratio
$(P/T)_{\pm}=\pm r_{\pm} e^{i\phi_{\pm}}/\sqrt{\bar\rho^2+\bar\eta^2}$
, where $(r_{\pm},\phi_{\pm})$ are pure strong interaction quantities
and $(\bar\rho,\bar\eta)$ are the perturbatively improved Wolfenstein
parameters\cite{Wolf}.
Neglecting the very small effects from electroweak penguin contributions
in our processes, one can express the penguin parameters $r_+ e^{i\phips}$
and $r_- e^{i\phims}$, respectively, in the form\cite{BBNS1,BBNSPV,Aleksan:2003qi}
\begin{eqnarray}\label{rqcd}
\frac{-(a^c_{V,4}+ r_A[b_{P,3}+b_{P,4}+b_{V,4}])}{
 a_{V,1}+a^u_{V,4}+ r_A[b_{V,1}+b_{P,3}+b_{P,4}+b_{V,4}]},\quad
\frac{-(a^c_{P,4}-r^\pi_\chi a^c_{P,6}+ 
k_A[b_{V,3}+b_{V,4}+b_{P,4}])}{
 a_{P,1}+a^u_{P,4}-r^\pi_\chi a^u_{P,6}+ 
k_A[b_{P,1}+b_{V,3}+b_{V,4}+b_{P,4}]}.
\end{eqnarray}
A recent analysis gives\cite{BS-preparation}:
\vspace{-0.5cm}
\begin{eqnarray}
\label{rphi}
r_{+} &=& 0.04 \pm 0.01, \,\,\,\qquad \phi_{+}= 0.18 \pm 0.27,\nn\\
r_{-} &=& 0.03 \pm 0.02, \,\,\,\qquad \phi_{-} = -0.02 \pm 0.42,\nn\\
r_t &=& 0.89 \pm 0.22,   \,\,\,\qquad \phi_{t} =  0.02 \pm 0.03,
\end{eqnarray}
where the error includes an estimate of potentially important
power corrections.
In order to obtain additional insight into the structure of
hadronic $B$-decay amplitudes, it will be also interesting to extract these 
quantities from other $B$-channels, via $SU(3)$-symmetry \cite{Gronau:2004tm}
, or using other methods.



Since the parameters $r_{\pm}$ and $\phi_t$ are small quantities, 
$\big(\mathit{"CP-violating"}~S, ~\mathit{"CP-conserving"}~\Delta S\big)$ and 
their corresponding rescaled quantities
$\big(\mathit{"CP-violating"}~\bar S, ~\mathit{"CP-conserving"}~\Delta\bar S
\big)$ \cite{Gronau:2004tm} are respectively well 
approximated, at the lowest order in $r_{\pm}$, 
by~\cite{BBNSPV,Gronau:2004tm}:
\vspace*{-0.2cm}
\begin{eqnarray}\nn
S &\dot=& {\rm fct_1}(r_{\pm},\phi_{\pm},\rt,\tau,\etab), \qquad
\Delta S \dot={\rm fct_2}(r_{\pm},\phi_{\pm},\rt,\phi_t,\tau,\etab),
\\
\bar S &\dot=& {\rm \bar fct_3}(r_{\pm},\phi_{\pm},\tau,\etab), \qquad\quad
\Delta \bar S \dot={\rm \bar fct_4}(r_{\pm},\phi_{\pm},\phi_t,\tau,\etab),
\label{SetabatLO}
\end{eqnarray}
where the observable $\tau\equiv\cot\beta$ has been introduced to relate 
the parameter $\bar\rho$ to $\bar\eta$, namely $\bar\rho = 1-\tau\, \bar\eta$,
to simplify our analysis in terms of only one CKM parameter. Thus, 
assuming that the parameter $\tau$ (or $\sin2\beta$) is known one could 
extract precisely $\etab$, as we will see below. Taking
$\tau=2.26\pm 0.22$, $\bar\eta=0.35\pm 0.04$ \cite{ckmfit}
and our penguin parameters results in (\ref{rphi}), we find from 
(\ref{SetabatLO}) that 
\vspace*{-1.cm}
\begin{eqnarray}\label{spred}
S&=& -0.26 
         \quad ^{+0.28}_{-0.19}\,(\tau)
         \quad ^{+0.47}_{-0.39}\,(\bar\eta)
         \quad\underbrace{ ^{-0.02[+0.01]}_{+0.02[-0.001]}}_{(r_+[\phi_+])}
         \quad\underbrace{ ^{+0.04[-0.01]}_{-0.04[-0.005]}}_{(r_-[\phi_-])}
         \quad ^{-0.005}_{+0.02}\,(\rt),\\
\bar{S}&=& -0.42 
         \quad ^{+0.27}_{-0.18}\,(\tau)
         \quad ^{+0.49}_{-0.39}\,(\bar\eta)
         \quad\underbrace{ ^{-0.04[+0.02]}_{+0.04[-0.01]}}_{(r_+[\phi_+])}
         \quad\underbrace{ ^{-0.01[-0.005]}_{+0.01[+0.01]}}_{(r_-[\phi_-])}.
\end{eqnarray}
\begin{figure}[t]
\psfrag{SS}{\small{$S$}}
\psfrag{etab}{\small{$\bar\eta$}}
\begin{center}
\epsfig{file=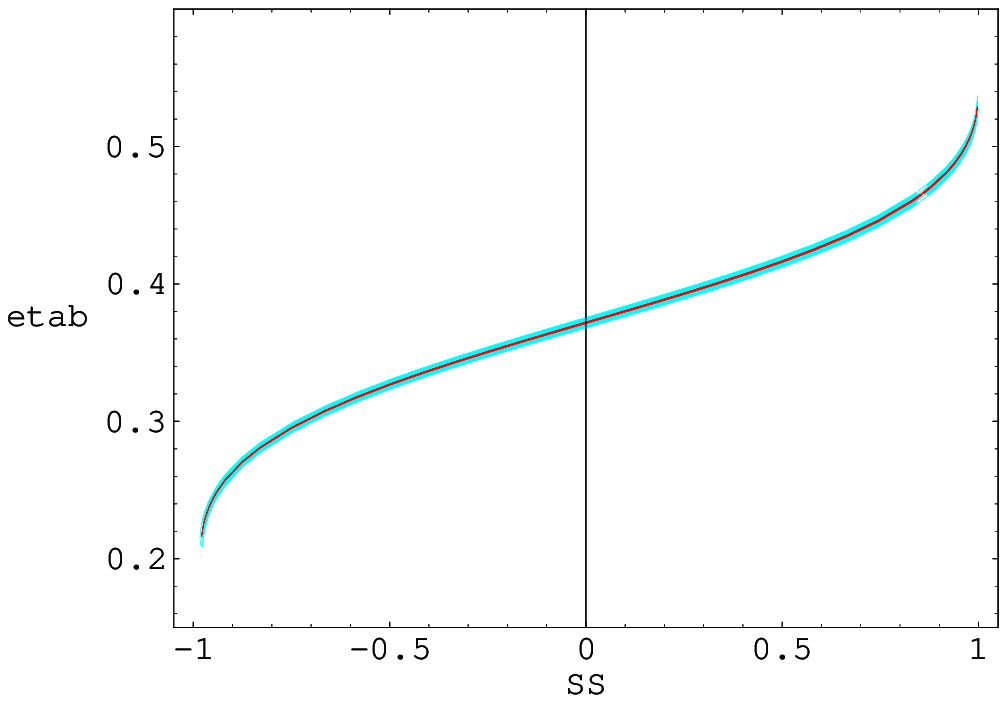,width=4cm,height=3cm}
\psfrag{SS}{\small{$S$}}
\psfrag{etab}{$$}
\epsfig{file=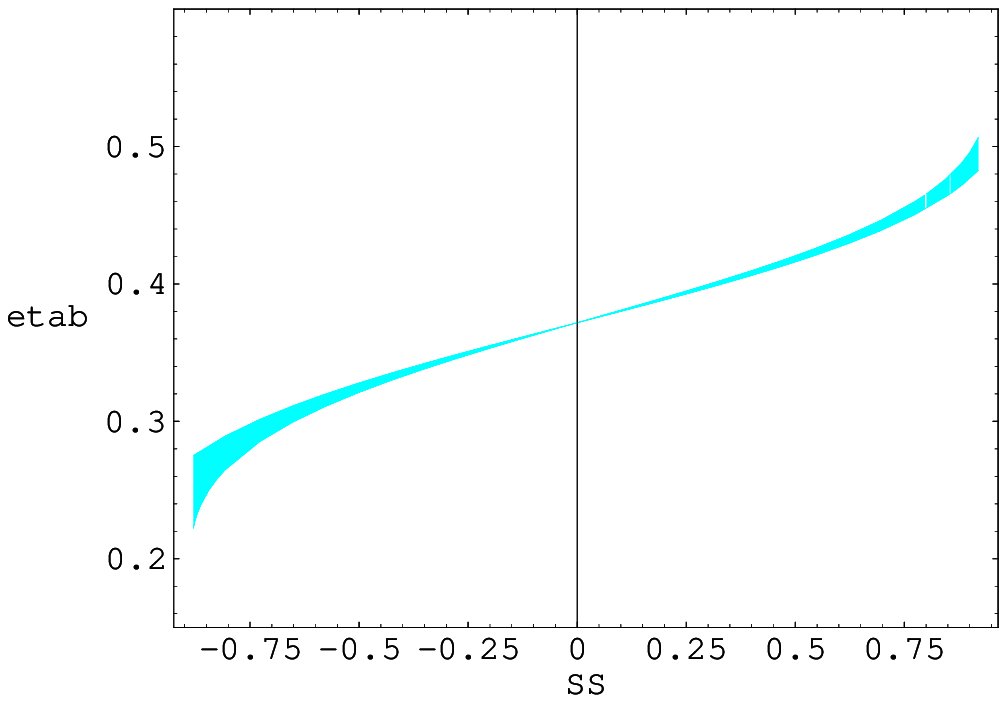,width=4cm,height=3cm}
\psfrag{SS}{$\bar S$}
\psfrag{etab}{$$}
\epsfig{file=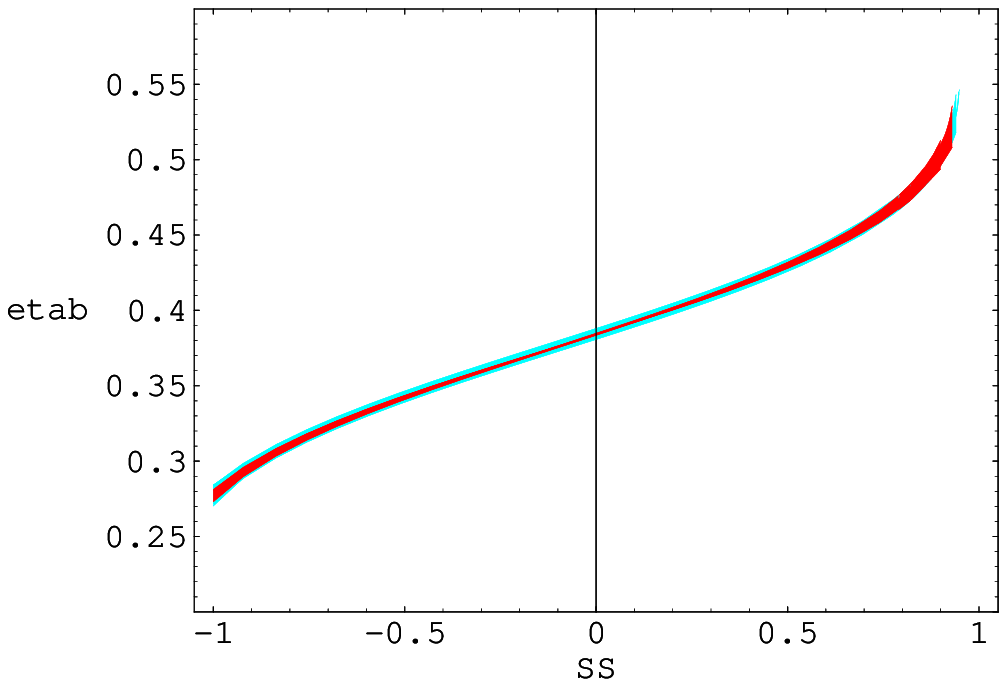,width=4cm,height=3cm}
\end{center}
\vspace{-0.7cm}
{\it{\caption{
CKM phase $\bar\eta$ as a function of the mixing-induced
$CP$ asymmetry $S$ or $\bar S$ in $B_d\to \rho^{\pm}\pi^{\mp}$ within the SM
for $\sin2\beta=0.739$.
The dark (light) band reflects the theoretical uncertainty
in the penguin phases $\phi_{\pm}$ (penguin amplitude $r_{\pm}$)
in the left and right plots, however in the middle one 
the band reflects the theoretical uncertainty
in the tree ratio $r_t$.    
\label{fig:etabs}
}}}
\end{figure}
%
%
We note that the sensitivity of $\tau$ or $\sin 2\beta$ in extracting 
$S$ is significant, however the dominant uncertainty is related 
to the $\bar\eta$ which for the purpose of predicting $S$ has 
been borrowed from a standard CKM fit \cite{ckmfit}. 
Distinctly, the large sensitivity of $S$ to $\bar\eta$ is 
analogous  to the fact that in turn $\bar\eta$ depends 
weakly on $S$.
Concerning $\bar S$, which is free from $r_t$ per definition,  
the impact of $\bar\eta$ and $\tau$ (or $\sin 2\beta$)
are significant, however less than in the case of $S$. 
%
In Fig.~\ref{fig:etabs}, we have plotted the CKM parameter 
$\etab$ as function of $S$, showing the $(r_{\pm},\phi_{\pm})$ (left-plot) 
and the $r_t$ (middle-plot) uncertainties.
We find that the 
sensitivity of $\bar\eta$ on the strong phase $\phi_{\pm}$ is rather mild
compared to the penguin amplitude $r_{\pm}$. This is not surprising, since the
 dependence on $\phi_{\pm}$ enters in  
$\bar\eta_{S}={\rm G}(r_{\pm},\phi_{\pm},\dt,\tau,S)$ only at second order.
As a cross check of our strategy one can extract $\etab$, 
independently of $\rt$, as function of $\bar S$, 
$\bar\eta_{\bar S}={\rm K}(r_{\pm},\phi_{\pm},\tau,\bar S)$.
In this case, however the sensitivity of $\bar\eta_{\bar S}$ on the strong 
phase $\phi_{\pm}$ is more pronounced than in $\bar\eta_{S}$, 
as shown in Fig.~\ref{fig:etabs} (right-plot). This 
is traced back to the fact that the strong phase 
$\phi_{\pm}$ in $\bar\eta_{\bar S}$, contrary to $\bar\eta_{S}$,   
enters at first order.
 
\begin{figure}[t]
\psfrag{phips}{\hspace*{0.2cm} \small{$\phi_+$}}
\psfrag{rps}{\small{$r_+$}}
\begin{center}
\epsfig{file=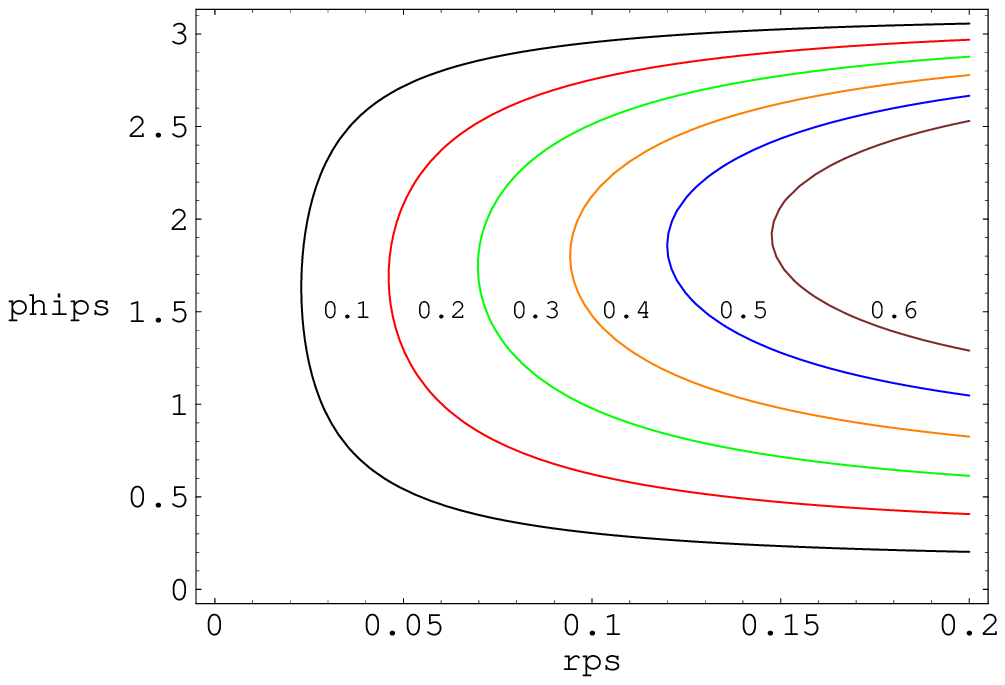,width=6cm,height=3cm}
\psfrag{phims}{\hspace*{0.2cm} \small{$\phi_-$}}
\psfrag{rms}{\small{$r_-$}}
\epsfig{file=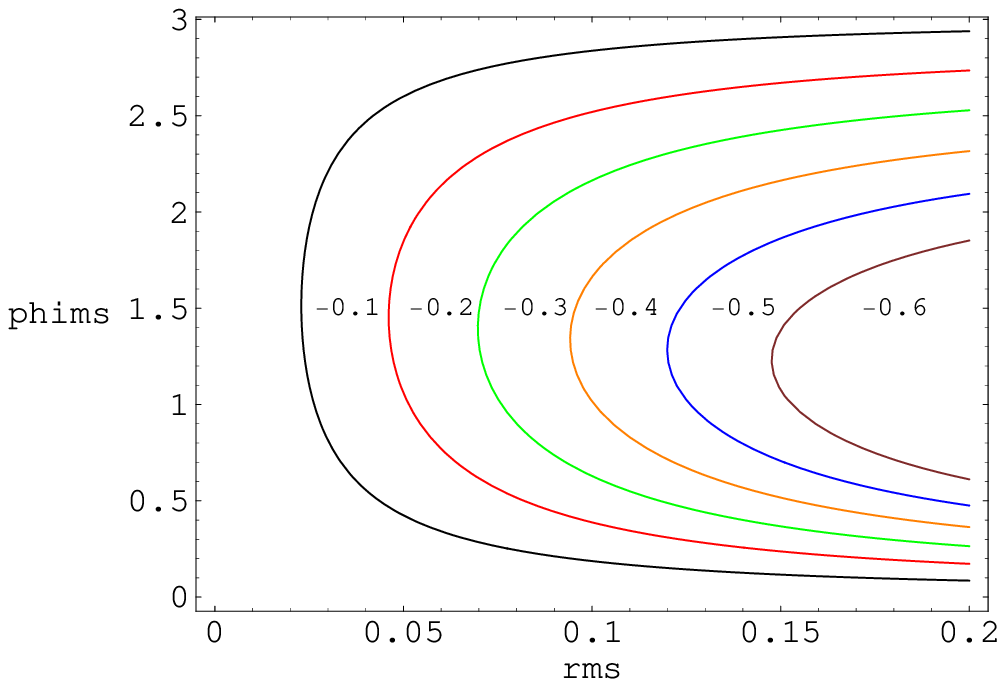,width=6cm,height=3cm}
\end{center}
\vspace{-0.7cm}
{\it{\caption{Contours of constant $C_{\pm}$ in the ($r_{\pm}$, 
$\phi_{\pm}$)-plane
for fixed $\bar\rho=0.20$ and $\bar\eta=0.35$.  \label{fig:cpm}}}}
\end{figure}

Moreover, the $B_d\to\rho^\pm\pi^\mp$ decays offer the possibility to explore 
the two individual direct CP asymmetries between $B^0(\Bbar)\to\rho^+\pi^-$ 
and $\Bbar(B^0)\to\rho^-\pi^+$ decay rates, namely $C_{\pm}$, 
For any given values of $r_{\pm}$ and $\phi_{\pm}$ a measurement of 
$C_{\pm}$ defines a curve in the ($\bar\rho$, $\bar\eta$)-plane,
%
as sketched in Fig. \ref{fig:cpm} where a model-independent correlation,
within the SM,
between the penguin parameters $r_{\pm}$ and $\phi_{\pm}$ for different 
values for $C_{\pm}$ is shown.
%
%
In the determination of $\bar\eta$ and $\bar\rho$ described
here discrete ambiguities do in principle arise. However, they can be excluded
by other information on the UT (for further details on these ambiguities, 
see \cite{BS}).


In \cite{BS1,BS}, it has been shown that the sensitivity of 
$\bar\eta$, in $B_d \to \pi^+ \pi^-$ modes, on the strong phase 
$\phi_{\pi\pi}$ is rather mild, since its dependence enters in $\bar\eta$ 
only at second order. Hence, using the lowest order result in $\phi_{\pi\pi}$
is most likely a very good approximation to the exact result 
(see \cite{BS1,BS}).
In an analogous manner, the same argument holds for $B_d \to \rho^{+}\rho^{-}$ 
channels. The corresponding equations are similar to those defined for 
$B_d \to \pi^+ \pi^-$ \big(see eqs. (36) and (37) in \cite{BS}\big).
Therefore, it is interesting to define the ratio \cite{BS-preparation}:
\vspace*{-0.2cm}
\begin{equation}\label{ratio}
\underbrace{\frac{\Srho(1+\tau \Spi -\sqrt{1-\Spi^2})}
{\Spi(1+\tau \Srho -\sqrt{1-\Srho^2})}}_{
{\mathcal R}^{\mathrm{exp}}}
=\underbrace{\frac{1+r_{\rho\rho}\cos\phi_{\rho\rho}}
{1+r_{\pi\pi} \cos\phi_{\pi\pi}}}_{{\mathcal R}^{\mathrm{the}}}.
\end{equation}
Note that the lhs of (\ref{ratio}), namely ${\mathcal R}^{\mathrm{exp}}$,
depends only on experimental observables, and is therefore a measurable
quantity, to be confronted with the theoretical informations encoded in the 
rhs of (\ref{ratio}) ${\mathcal R}^{\mathrm{the}}$.
Since $r_{\pi\pi}$ $\sim r_{\rho\rho}\sim 0.1$ are small, we expand the ratio 
${\mathcal R}^{\mathrm{the}}$ in these two parameters to get to lowest order 
${\mathcal R}^{\mathrm{the}}
\dot=1+{\mathrm{Re}}[r_{\rho\rho}~e^{i\phi_{\rho\rho}}-r_{\pi\pi}~e^{i\phi_{\pi\pi}}],$
where in QCDF 
$r_{\rho\rho}~e^{i\phi_{\rho\rho}}-r_{\pi\pi}~e^{i\phi_{\pi\pi}}
\approx \frac{r^{\pi}_{\chi} a_6^c}{a_1}$
is very sensitive to the chirally enhanced terms $r^{\pi}_{\chi}$. Using QCDF, 
we find
$\frac{r^{\pi}_{\chi} a_6^c}{a_1}=-0.06\pm 0.014$, leading to
$\mathcal{R}^{\mathrm{the}}_{\mathrm{QCDF}}=0.94\pm{0.014}$ \cite{BS-preparation}.
On the other hand, using the most recent experimental data reported either
by BaBar or Belle concerning $(S_{\pi\pi}, S_{\rho\rho}, \tau)$ \cite{HFAG}, 
we find
$\mathcal{R}^{\mathrm{exp}}_{\mathrm{BaBar}}=1.01\pm{0.11}$ and 
$\mathcal{R}^{\mathrm{exp}}_{\mathrm{Belle}}=0.81\pm{0.11}$.
Although these two values suffer from large uncertainties, their central 
values agree quite remarkably with the QCDF one given above.

%



%
We proposed strategies to extract
information on weak phases from CP violation
observables in $B_d\to\rho^{\mp}\pi^{\pm}$ decays even in the presence
of hadronic contributions related to penguin amplitudes.
Assuming knowledge of the penguin pollution, an efficient use 
of mixing-induced CP violation in $B_d\to\rho^{\mp}\pi^{\pm}$
decays, can be made by combining it with the corresponding
observable from $B_d\to J/\psi K_S$, $\sin 2\beta$, to obtain the 
UT parameters $(\bar\rho,\bar\eta)$.
The sensitivity on the hadronic quantities is discussed.
In particular, there are no first-order corrections in $\phi_{\pm}$.
Moreover, we found that the dependence on $r_{\pm}$ in 
$B_d\to \rho^{\pm} \pi^{\mp}$ is less pronounced
than in the $B_d \to \pi^+ \pi^-$ case, since $r_{\pm}\approx r_{\pi\pi}/3$, 
leading to an important simplification in our analysis.
Finally, a new experimental test of the factorization ansatz is
presented  using $B_d \to \pi^+ \pi^-$ and $B_d \to \rho^{\pm}
\rho^{\mp}$ modes.
%

\vfill\eject

\end{document}